\documentclass[12pt]{article}
\usepackage{epsfig, epsf, graphicx, subfigure, epstopdf}
\usepackage{pstricks, pst-node, psfrag}
\usepackage{amssymb,amsmath,bm}
\usepackage{verbatim,enumerate}
\usepackage{rotating, lscape}
\usepackage{setspace}
\usepackage{multirow}
\usepackage{color}
\usepackage[normalem]{ulem}
\usepackage{lineno}

\usepackage{natbib}

\def\boxit#1{\vbox{\hrule\hbox{\vrule\kern6pt
          \vbox{\kern6pt#1\kern6pt}\kern6pt\vrule}\hrule}}

\def\bse{\begin{eqnarray*}}
\def\ese{\end{eqnarray*}}
\def\be{\begin{eqnarray}}
\def\ee{\end{eqnarray}}
\def\bq{\begin{equation}}
\def\eq{\end{equation}}
\def\bse{\begin{eqnarray*}}
\def\ese{\end{eqnarray*}}

\newcommand{\bA}{\mathbf{A}}
\newcommand{\bB}{\mathbf{B}}

\newcommand{\bI}{\mathbf{I}}
\newcommand{\bR}{\mathbf{R}}

\newcommand{\bU}{\mathbf{U}}
\newcommand{\bW}{\mathbf{W}}
\newcommand{\bX}{\mathbf{X}}
\newcommand{\bY}{\mathbf{Y}}
\newcommand{\bZ}{\mathbf{Z}}

\newcommand{\bz}{\mathbf{z}}
\newcommand{\bmu}{\boldsymbol{\mu}}

\newcommand{\bomega}{\boldsymbol{\omega}}
\newcommand{\bOmega}{\boldsymbol{\Omega}}
\newcommand{\bgamma}{\boldsymbol{\gamma}}
\newcommand{\bxi}{\boldsymbol{\xi}}
\newcommand{\bdelta}{\boldsymbol{\delta}}

\newcommand{\bvepsilon}{\boldsymbol{\varepsilon}}
\newcommand{\balpha}{\boldsymbol{\alpha}}
\newcommand{\bbeta}{\boldsymbol{\beta}}

\newcommand{\bSigma}{\boldsymbol{\Sigma}}

\newcommand{\0}{\mathbf{0}}

\begin{document}

\thispagestyle{empty} \baselineskip=28pt \vskip 5mm
\begin{center} {\huge{\bf A Non-Gaussian Spatio-Temporal Model for Daily Wind Speeds
Based on a Multivariate Skew-$t$ Distribution}}
\end{center}

\baselineskip=12pt \vskip 10mm

\begin{center}\large
Felipe Tagle\footnote[1]{\baselineskip=10pt Department of Applied and Computational Mathematics and Statistics, University of Notre Dame, Notre Dame, IN 46556, United States.  \\ E-mail: ftagleso@nd.edu,  scastruc@nd.edu}, Stefano Castruccio$^{1}$, Paola Crippa\footnote[2]{\baselineskip=10pt Department of Civil \& Environmental Engineering \& Earth Sciences, University of Notre Dame, Notre Dame, IN 46556, United States. \\ E-mail: pcrippa@nd.edu} and Marc G.~Genton\footnote[3]{\baselineskip=10pt Statistics Program, King Abdullah University of Science and Technology, Thuwal 23955-6900, Saudi Arabia. \\ E-mail: marc.genton@kaust.edu.sa \\
This publication is based upon work supported by the King Abdullah University of Science and Technology (KAUST) Office of Sponsored Research (OSR) under Award No: OSR-2015-CRG4-2640.}\\
\end{center}

\baselineskip=17pt \vskip 10mm \centerline{\today} \vskip 10mm

\begin{center}
{\large{\bf Abstract}}
\end{center}
\baselineskip=17pt

Facing increasing domestic energy consumption from population growth and industrialization, Saudi Arabia is aiming to reduce its reliance on fossil fuels and to broaden its energy mix by expanding investment in renewable energy sources, including wind energy. A preliminary task in the development of wind energy infrastructure is the assessment of wind energy potential, a key aspect of which is the characterization of its spatio-temporal behavior. In this study we examine the impact of internal climate variability on seasonal wind power density fluctuations over Saudi Arabia using 30 simulations from the Large Ensemble Project (LENS) developed at the National Center for Atmospheric Research. Furthermore, a spatio-temporal model for daily wind speed is proposed with neighbor-based cross-temporal dependence, and a multivariate skew-$t$ distribution to capture the spatial patterns of higher order moments. The model can be used to generate synthetic time series over the entire spatial domain that adequately reproduce the internal variability of the LENS dataset.

\par\vfill\noindent
{\bf Some key words:}  daily wind speed, skew-$t$ distribution, internal climate variability, wind power density
\par\medskip\noindent
{\bf Short title}: Non-Gaussian Spatio-Temporal Model for Daily Wind Speeds

\clearpage\pagebreak\newpage \pagenumbering{arabic}
\baselineskip=26.5pt

%
\section{Introduction}\label{sec:intro}


Wind energy has become an important component of energy portfolios for many developed and developing nations worldwide. This trend is largely driven by technological advances that enable more efficient harnessing of available wind energy and its integration into energy transmission networks, but is also due to growing recognition of the importance of renewable energy sources in climate change mitigation strategies \citep[][and references therein]{bruckner2014climate,zhu2012short}. For instance, Denmark relies on wind energy for nearly 40\% of its domestic energy consumption, and this number is expected to rise to 50\% by 2020 \citep{dea2016}. In 2015, China added 33 GW of new installed capacity, to reach a total capacity of 148 GW, representing an increase of 29\% \citep{wwea2016}. These and other examples provide convincing evidence that wind energy offers a viable alternative to traditional fossil fuel-based energy sources, and are paving the way for other nations to exploit this renewable and clean energy resource. 

Saudia Arabia, with its vast oil reserves, has a long tradition of relying solely on fossil fuels for its energy needs. However, faced with rising energy demands due to population growth and industrialization, it is actively seeking to diversify its energy mix by expanding its renewable energy portfolio to 54 GW by 2032, of which 9 GW will come from wind power \citep{kacare2012}. Due to the inherent variability and limited predictability of the wind resource, assessing its spatial and temporal characteristics is a critical step in the development of wind energy infrastructure. Several studies have sought to quantify the wind potential over Saudi Arabia \citep[e.g.,][]{rehman2004assessment, rehman2007wind,shaahid2014potential}, however, these have mostly focused on just a few locations, largely due to the country's sparse observational network with varying record lengths. A recent study by \citet{yip2016wind} used a gridded reanalysis dataset with a multi-decadal record period to provide a more comprehensive assessment, evaluating the abundance of the wind resource over the Arabian Peninsula. The authors also investigated its variability and intermittency, as these pose the greatest challenges to the integration of wind energy into existing power grid systems. In this study, as part of an ongoing collaborative effort with the National Center for Atmospheric Research (NCAR), we focus on the spatio-temporal characteristics of this resource over Saudi Arabia. Wind speeds fluctuate over a wide range of frequencies, with those of the order of days governed by general weather patterns, while shorter frequencies are driven by local effects and turbulence \citep{pinson2013wind}. Our focus is on very low frequencies, where long-term wind trends and the effects of human activities may potentially play a role. In particular, this is the first study that investigates the sensitivity of wind energy potential to internal climate variability, that is, the variability that is intrinsic to the climate system due to the complex interactions between its components at various temporal and spatial scales. Only recently has this source of variability been properly identified in the geophysical community \citep[i.e.,][]{deser2012uncertainty,deser2014projecting}, and its impact is just beginning to be understood. Because of the decadal timescales at which this source of variability operates, the characterization of its impact usually relies on simulations from Earth System Models (ESMs), which, in addition to solving the atmospheric governing equations, include processes from other components of the climate system, namely, land, ocean and sea ice; and so are well-suited to represent the complex interactions that occur at such timescales. In this work we use a collection of simulations based on the Community Earth System Model (CESM), developed at NCAR under the Large Ensemble Project (LENS) that are specifically designed to isolate the effects of internal climate variability. These simulations and ESM simulations in general, however, are computationally expensive to run, and many simulations are necessary to properly quantify the effects of natural variability. An alternative to simulating atmospheric processes using these models is provided by stochastic weather generators \citep{wilks1999weather, ailliot2015stochastic}, which are statistical models designed to generate synthetic sequences of meteorological variables whose statistical properties closely resemble those of the observational datasets on which they are trained. We develop one such model designed to capture the natural variability of daily wind speed in Saudi Arabia, that may be used, among other applications, to assess future wind energy potential under different climate scenarios. 

Stochastic weather generators are widely used in applications ranging from agricultural models to climate impact studies. For instance, they are commonly used in the energy industry to generate short-temporal frequency simulations of wind data to assist utilities in grid integration studies, where these and other components of an electrical grid are simulated to assess general performance measures and determine best practices \citep{archer2017challenge}. In \citet{ailliot2012markov}, a model of this kind was used to generate wind time series at different meteorological stations near potential wind farm sites to assess various quantities related to wind power production. Most of these stochastic wind generators have focused on single-site models, and typically use Markov chains to describe the temporal dependence on different wind regimes \citep{ailliot2006autoregressive}. Recently, focus has shifted towards a multisite framework, which poses the challenge of having to explicitly account for the complex cross-correlation patterns among neighboring sites \citep{bessac2015gaussian, hering2015markov, bessac2016comparison}. Thus, the number of sites under consideration is usually limited, because the quality of fit quickly deteriorates as the quantity increases \citep{hering2015markov}. The model presented here, however, with the approximately $1^\circ$ horizontal resolution of the LENS dataset, is applied to a 149-point spatial domain which, together with a temporal coverage spanning more than 80 years, results in a spatio-temporal domain of nearly 5 million data points. To emphasize the different framework of this statistical model, consistently with \cite{jeo18,jeo19}, we denote this as a \textit{stochastic generator}, and in particular in this application, a \textit{stochastic wind generator}. While stochastic generators rely on a stochastic approximation of climate model output, they are also fundamentally different from an emulator in that they are not used to explore the input parameter space and perform sensitivity analysis.

Environmental data often exhibit departures from Gaussianity, such as skewness and heavy tails. Traditionally, to continue exploiting the appealing properties of the normal distribution and the well-developed theory of Gaussian processes, a transformation is usually applied to the data; e.g., a square root transformation \citep{gneiting2002nonseparable}, a power transformation \citep{ailliot2015non, bessac2016comparison} or a Gaussian copula \citep{hering2015markov}. Only recently have there been studies that exploit the flexibility of skew-elliptical distributions, of which the skew-$t$ \citep{azzalini2003distributions} and the skew-normal \citep{azzalini2005skew} distributions are special cases that directly address the skewness and excess kurtosis that are commonly found in wind data. For example, \citet{hering2010powering} developed a short-term wind speed forecasting model using a bivariate skew-$t$ distribution, while \citet{flecher2010stochastic} used a closed skew-normal distribution \citep{gonzalez2004closed} as part of a multivariate weather generator. This family of distributions represents an extension of the normal model and retains several of its convenient properties, such as closure under conditioning and marginalization (see \citet{genton2004skew} or
\citet{azzalini2014skew} for an overview of the theory and applications). In this paper we propose a stochastic wind generator that leverages the flexibility of the skew-t distribution to capture the non-Gaussian features that are common to wind-speed time series. In particular, we consider a vector-autoregressive process to capture the temporal auto- and cross-correlations, coupled with independent realizations from a multivariate skew-$t$ distribution as a model for the spatial residuals. Given that our interest lies in generating spatial replicates at locations and support that coincide with those of the training dataset, we consider the skew-$t$ distribution only in its multivariate form. An extension to a spatial process \citep[e.g.,][]{morris2017space,beranger2017models} would have been necessary had the scope of this study involved an application of spatial interpolation, but it is not contemplated here.


The remainder of the paper is organized as follows. Section \ref{sec:data} describes the daily wind speed data used for the training and validation of the spatio-temporal model described in Section \ref{sec:model}. Section \ref{sec:saudi} provides an assessment of the temporal variability of the wind energy resource in Saudi Arabia, and Section \ref{sec:disc&conc} offers a discussion and conclusion.

\section{Wind Data}\label{sec:data}

We consider daily wind speed in Saudi Arabia from the publicly available LENS dataset developed at NCAR \citep{kay2015community}. The ensemble consists of 30 fully-coupled simulations, based on CESM version 1, with the Community Atmosphere Model (CAM), version 5, run at approximately $1^\circ$ horizontal resolution in all model components. The simulations span from the year 1920 to 2100, with radiative forcing following the CMIP5 protocol; namely, historical forcing from 1920 to 2005, and the representative concentration pathway 8.5 (RCP8.5) from 2006 to 2100. Each simulation represents a unique climate trajectory, generated by introducing small round-off level differences into their initial atmospheric conditions. For this study we use the historical segment (1920-2005) of 30 ensemble members over Saudi Arabia, bounded roughly by $15$-$32^\circ$N and $35$-$55^\circ$E, which at the noted horizontal resolution corresponds to $N$=149 points, in the temporal dimension $T$=31,390 points (the model does not account for leap years), resulting in nearly 5 million data points.


\begin{figure}[!bp]\centering
\includegraphics[width=\textwidth]{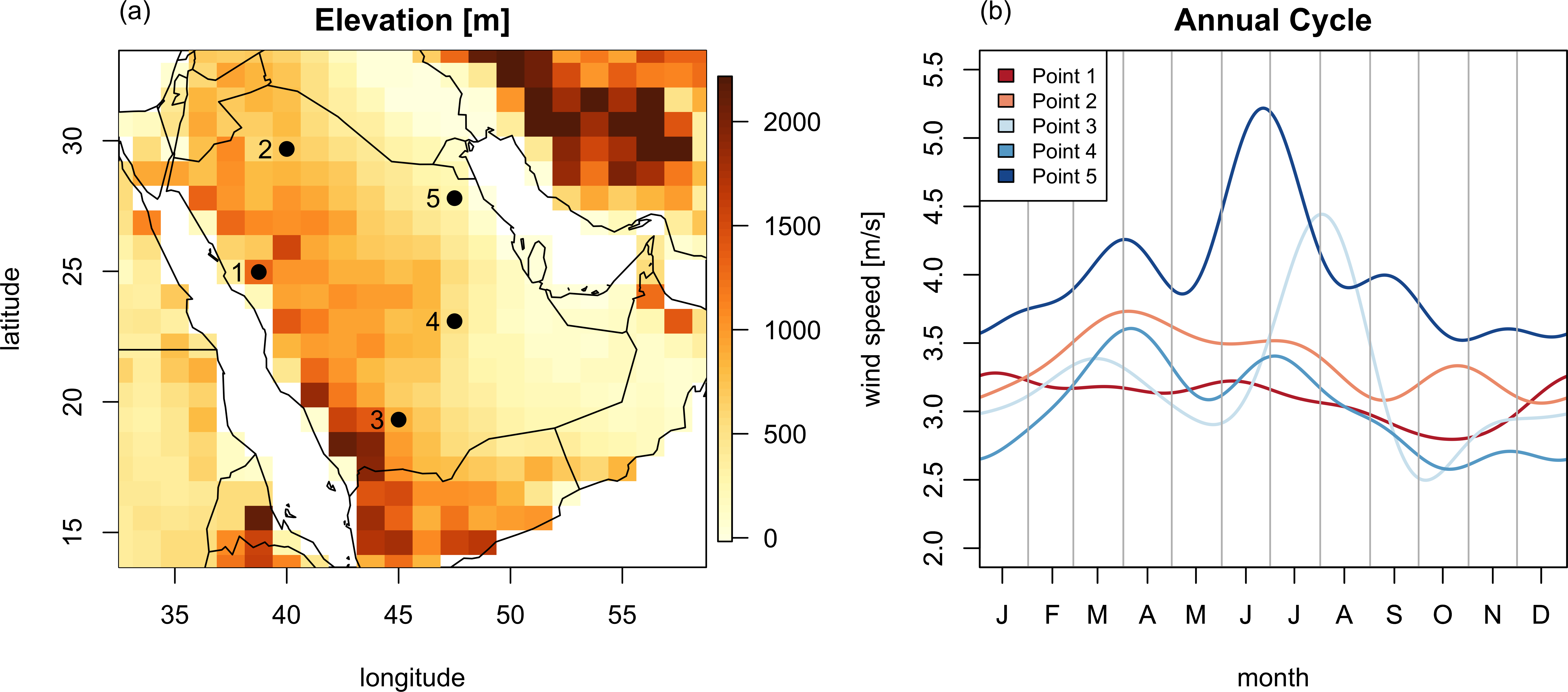}
\caption{(a) Selection of points over Saudi Arabia with colors showing the elevation in meters, and their respective (b) daily wind speed annual cycles, computed as the mean annual cycle from the LENS. Wind speed data is based on the 1920-2005 output.}
\label{fig:AnnCycle}
\end{figure}

In order to illustrate the diverse wind regimes across Saudi Arabia, we examine the various shapes of the annual cycle, as represented by the LENS ensemble mean. Figure \ref{fig:AnnCycle} displays the annual cycles for five locations that broadly capture the variety of regimes. The seasonal patterns can be interpreted with the aid of the mean monthly wind fields (Fig.~S.1 in the supplementary material), constructed from the lowest level zonal and meridional winds (UBOT and VBOT, respectively) obtained from the LENS database. The Arabian Peninsula lies within the trade-wind belt (Hadley cell) of the Northern Hemisphere. During winter, westerlies from the Mediterranean travel southward towards the Persian Gulf (Shamal trade winds), and turn south and southwest through the Rub al-Khali Desert toward Yemen. But beginning in spring, the northerly winds intensify, and the onset of the monsoon circulation from the Indian Ocean induces a southeasterly flow along the southeastern part of the peninsula and the Arabian Sea, into the Rub al-Khali Desert. As a result, the mild easterly flow over Point 3, located east of the Asir Mountains, becomes strongly westerly during the summer, explaining the maximum daily wind speed during this season. Moreover, the low speeds in March and September, correspond to the transitions between easterly and westerly flow, characterized by strong wind convergence. Despite their distance, Points 2 and 4 display a similar pattern, with dual peaks arising in early spring and the middle of summer, largely due to the intensification of the northerly winds during the monsoon season. The temporal pattern over Point 5, close to the eastern coast, most clearly reflects the second intensification of the Shamal trade winds during May and June. The mountainous region on the eastern coast of the peninsula, as represented by Point 1, displays a dampened annual cycle, reflecting the effect of the complex orography.

\section{Model Description}\label{sec:model}

In this section we describe the proposed spatio-temporal model, and discuss inferential aspects, highlighting several difficulties and the approach undertaken to overcome them. Finally, we validate the model along several metrics that focus on low-frequency aspects of the wind-speed time series.
 
\subsection{Spatio-temporal model}

Let $\mathbf{W}_{t,r} = (W_{1,t,r},\ldots,W_{N,t,r})^\top$ denote the $N$-vector of daily wind speeds over the domain at day $t$, in realization $r$. We first consider a standardization of the form $(W_{i,t,r} - \mu_{i,t})/\sigma_{i,t}$, where $\mu_{i,t}$ refers to a seasonal effect and $\sigma_{i,t}$ the seasonal fluctuation in the standard deviation of the residuals, both indexed by time $t$ and location $i$. The former is estimated by regressing the mean of each calendar day of the time series of daily wind speed at each gridpoint on a small set of harmonics, ranging from frequencies of 1 to 5 cycles per year. Similarly, the latter is obtained by fitting the same range of harmonics to the standard deviation of each calendar day of the residual time series at each gridpoint. We make the assumption that both standardization terms are common to all ensemble members, to be consistent with the design of the LENS. Despite the realizations having a time span of over 80 years, we did not find evidence of temporal non-stationarity in the estimates (see the diagnostics in Figs.~S.2, S.3, S.4 and S.5). For ease of notation, we assume henceforth that $\mathbf{W}_{t,r}$ and its component terms have undergone said standardization. For the residuals, we propose a model of the form \vspace{.08in}
\begin{equation}\label{eq:Model}
\bW_{t,r} = \bA_1\bW_{t-1,r} + \bA_2 \bW_{t-2,r} + \bvepsilon_{t,r} \quad t=3,\ldots,T,
\end{equation}
where $\bA_k$, $k=1,2$, are $N \times N$ coefficient matrices and $\bvepsilon_{t,r}$ corresponds to the time $t$ innovations of the $r$-th realization, for which $\text{E}(\bvepsilon_{t,r}) = \0$, and $\text{E}(\bvepsilon_{t,r} \bvepsilon_{t,r}^\top) = \bSigma_{\bvepsilon}$. We assume that the coefficient matrices are identical across the ensemble, while the distinct trajectories of each realization are the result of i.i.d. innovations. This is in agreement with the construction of the LENS and atmospheric flow more generally; the initial condition memory is lost within a few weeks, after which each ensemble member evolves chaotically as if driven by random atmospheric fluctuations \citep{kay2015community}. In order to assess whether fitting the model to a few realizations is sufficient to capture the variability across the LENS, the model in \eqref{eq:Model} is fit to the time series belonging to three members under the above assumption of i.i.d. realizations. These three are chosen arbitrarily among the 30 realizations, replicating a scenario where only 3 members are available. However, the results are robust to number of members of the training set (see Fig.~S.6 and its associated section in the supplementary material).
 
The temporal dependence in wind time series at both hourly and daily time scales is usually well represented in the univariate case by an autoregressive model of order 2 \citep[e.g.,][]{brown1984time,haslett1989space,ailliot2006autoregressive} and in the multi-site setting, by its multivariate counterpart, the vector-autoregressive (VAR) process of order 2 \citep[e.g.,][]{hering2015markov,bessac2016comparison}. An obvious difficulty that emerges in the latter context is the estimation of the autoregressive matrices as the dimension of the multivariate process grows. Parametric models for off-diagonal elements based on the distance between sites or shrinkage methods for reducing elements to zero have been proposed to address this issue \citep{monbet2017sparse}. \citet{schweinberger2017high} investigates the $p \gg N$ case, where $p$ is the number of parameters and $N$ is the number of observations, under the assumption that informative spatial structure is available. Here the gridded nature of the spatial domain naturally suggests a nearest-neighbor specification. Consider the following notation: $\bW  = (\bW_3,\ldots,\bW_T)$, $\bB   = (\bA_1,\bA_2)$,  $\bZ_t  = \text{vec}(\bW_{t},\bW_{t-1})$, $\bZ  = (\bZ_2, \ldots,\bZ_{T-1})$ and $\bU  = (\bvepsilon_3,\ldots,\bvepsilon_T)$. Then the VAR(2) process in \eqref{eq:Model} can be expressed compactly as 
\[ \bW = \bB\bZ + \bU \]
or equivalently as $\bW^* =(\bZ^\top \otimes \bI_N)\bbeta + \bU^*$, which results from applying the vec operator to both sides of the equation, so that $\bW^*= \text{vec}(\bW)$, $\bbeta = \text{vec}(\bB)$ and $\bU^* = \text{vec}(\bU)$. The nearest-neighbor specification amounts to imposing zero restrictions on the non-neighbor coefficients, which correspond to linear constraints of the form $\bbeta = \bR \bgamma$, where $\bR$ is a $(2N^2) \times M$ matrix of rank $M$ that encodes the $A_{i,j,k} = 0$ restrictions, for $i,j = 1,\ldots,N$, $k=1,2$. Here, $\bgamma$ is an unrestricted $M \times 1$ vector of unknown parameters, representing the non-zero elements of $\bB$ and $M$ the number of such elements; see Chp.~5 in \citet{lutkepohl2005new} for details. The generalized least squares (GLS) estimator of $\bgamma$ is given by
\begin{equation}\label{GLS}
\hat{\bgamma} =[\bR^\top (\bZ\bZ^\top\otimes\bSigma_{\bvepsilon}^{-1})\bR]^{-1} \bR^\top (\bZ \otimes \bSigma_{\bvepsilon}^{-1}),
\end{equation}
where $\bSigma_{\bvepsilon}$ denotes the covariance matrix of $\bvepsilon_t$. The residuals arising from the GLS estimation of $\bgamma$, among several nearest-neighbor configurations, did not adequately capture the autoregressive structure, in comparison with the traditional OLS approach that replaces $\bSigma_{\bU^*}$ in \eqref{GLS} with $\bI_N$. The GLS approach attempts to account for the second order structure in the residuals through the autoregressive parameters, which is undesirable in this case as we will model that structure in a subsequent stage. We therefore opted to proceed with the OLS approach. The stability condition of the VAR(2) process is checked in the usual manner, verifying that all of the eigenvalues of the autoregressive matrix of the compact form have modulus less than one.

\begin{figure}[htbp]\centering
\includegraphics[width=\textwidth]{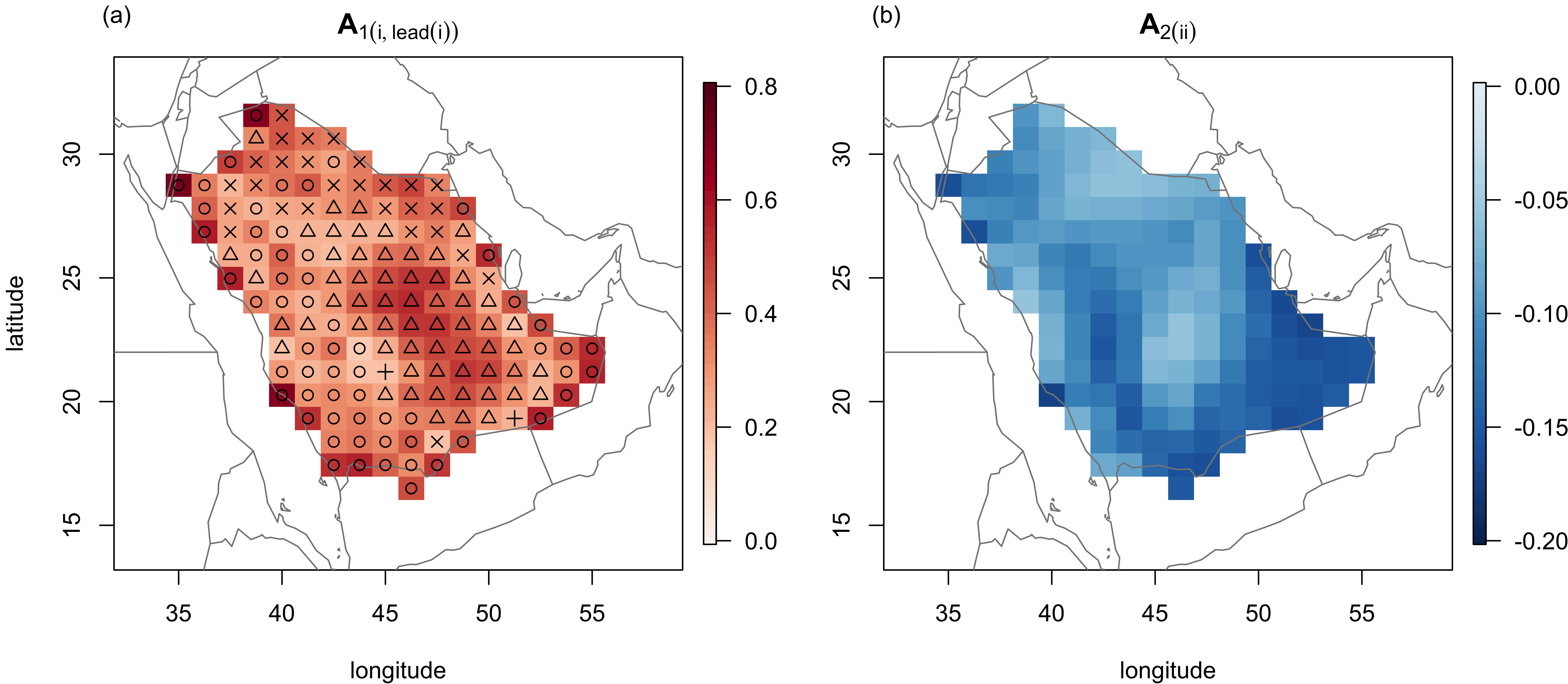}
\caption{Estimates for the (a) leading terms in $\bA_1$ and (b) the diagonal elements of $\bA_2$; all statistically significant at 1\%, withstanding an adjustment for multiplicity using the false discovery rate \citep{benjamini1995controlling}. Symbols in (a) indicate the neighbor corresponding to the leading term: circle, itself; triangle point-up, North; triangle point-down, South; slanted cross, West; cross, East.} 
\label{fig:AR2Coefs}
\end{figure}

We find that a first-order stencil neighborhood scheme (North, West, South, East) for $\bA_1$ and a diagonal form for $\bA_2$ adequately represents the temporal and cross-temporal structure (Fig.~S.7). Figure \ref{fig:AR2Coefs} displays the estimates for the leading and diagonal terms in $\bA_1$ and $\bA_2$, respectively. Symbols in Fig.~\ref{fig:AR2Coefs}(a) denote neighbor to which the value of the leading term corresponds. The estimates in $\bA_1$ are consistently positive across the entire domain, with values exceeding 0.7 along the boundaries. The leading terms in the northern part of the country are generally from the West neighbor, whereas those along the center belong to the North neighbor, in agreement with the predominant wind patterns in Fig.~S.1. Those of $\bA_2$, however, are consistently negative, and even more so in the Eastern Province.

\begin{figure}[htbp]\centering
\includegraphics[width=\textwidth]{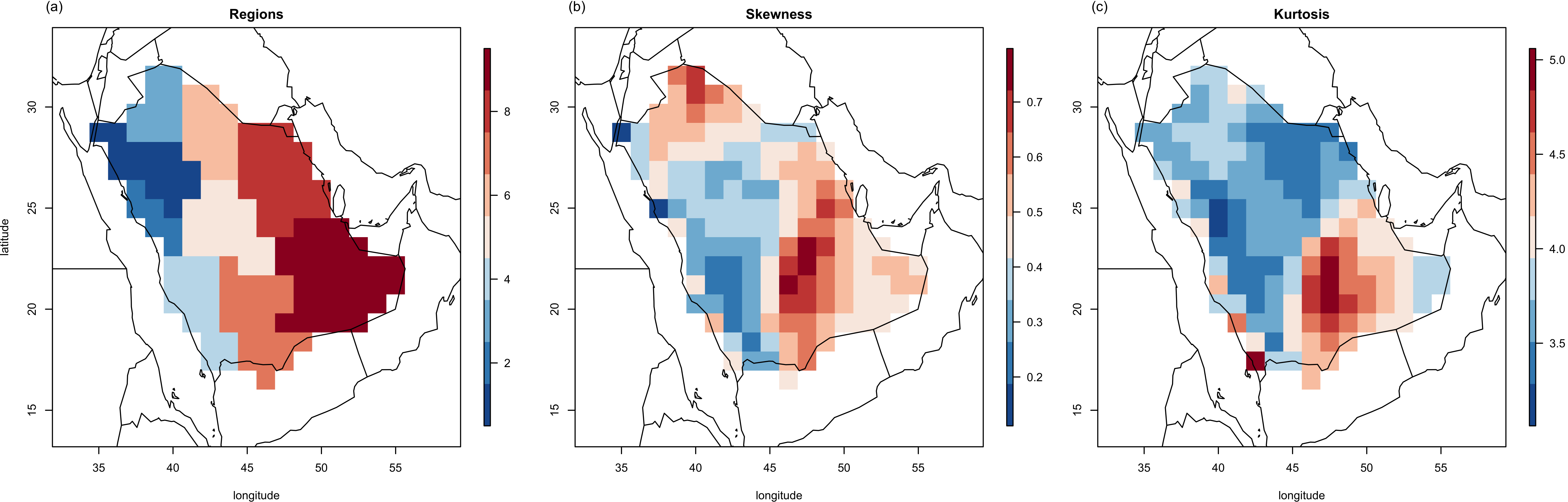}
\caption{(a) Nine regions used in the fitting of the innovation $\bvepsilon_t$, defined using Ward's method, an agglomerative method that minimizes the sum of squared Euclidean distances between the points and the centroids of their respective cluster. Here each point corresponds to the $N$-dimensional vector of rescaled VAR(2) residuals. The (b) skewness and (c) kurtosis coefficients have been added as visual guides to help explain the regional shapes.}
\label{fig:AR2Clusters}
\end{figure}

Given a domain of this size, with its wide range of wind regimes, the assumption of stationarity for a parametric form of the associated correlation matrix would be inappropriate. Several approaches have been advocated in the statistical literature to deal with non-stationary spatial covariance structures, beginning with the seminal work of \citet{sampson1992nonparametric} and their spatial deformation approach. More recently, other approaches based on spatially-weighted combinations of stationary spatial covariance functions \citep{fuentes2001high,fuentes2001new} and process convolutions \citep{higdon1998process,paciorek2006spatial} have received particular attention (see \citet{sampson2010constructions} for a review). Here we partition the spatial domain into regions where the assumption of stationarity is plausible; i.e., we divide the vector $\bvepsilon_t$ (henceforth we drop the subscript $r$ for convenience) into $N_c$ subvectors, $\bvepsilon_t = (\bvepsilon^\top_{1,t}, \ldots,\bvepsilon^\top_{N_c,t})^\top$, where each $\bvepsilon_{c,t}$, composed of $d_c$ number of gridpoints, is designed to capture regional features. We use Ward's hierarchical clustering method to partition the domain, as it tends to produce clusters of approximately equal size \citep{everitt2011clusters}. Nine regions (Fig.~\ref{fig:AR2Clusters}) result in cluster sizes that are suitable for parameter estimation. The clustering of spatial locations has a long history in environmental applications \citep{wilks2011statistical}. Recently, \citet{lorente2015characterization} used a combination of hierarchical and non-hierarchical techniques to cluster wind station data collected in the Iberian Peninsula. In the Markov regime-switching literature, clustering is often performed along the temporal dimension to identify different weather regimes \citep{kazor2015assessing}. 

Each $\bvepsilon_{c,t}$ is assumed to follow an independent multivariate skew-$t$ distribution, based on the formulation of \citet{azzalini2003distributions} which has an analogous derivation to the familiar Student-$t$ distribution, as opposed to others such as that of \citet{branco2001general}; see \citet{azzalini2016nomenclature} for a comparison of these formulations. A $d$-dimensional random variable $\bZ$ follows a multivariate skew-$t$ distribution, denoted by $ST_{d}(\bxi, \bOmega,\balpha, \nu)$, with  $\bxi$ being a location $d$-vector, $\bOmega$ a $d \times d$ scale matrix, $\balpha$ a $d$-vector skewness parameter, and $\nu$ being the degrees of freedom, if it has a probability density of the form 
\begin{equation}\label{eq:densT}
f(\bz) = 2t_d(\bz -\bxi; \bOmega,\nu) T\left(\balpha^\top \bomega^{-1}(\bz - \bxi) \sqrt{\frac{ \nu + d}{\nu + Q(\bz)}}; \nu + d \right),
\end{equation}
with $Q(\bz) = (\bz -\bxi)^\top \bOmega (\bz -\bxi)$, $\bomega = \text{diag}(\omega_1,\ldots,\omega_d)>0$, $\bOmega = \bomega \bar{\bOmega} \bomega$, $\bar{\bOmega}$ a $d \times d$ correlation matrix, $t_d(\cdot; \bOmega, \nu)$ is the probability density of the $d$-dimensional Student-$t$ distribution, and $T(\cdot; \nu)$ denotes the cumulative distribution function of a univariate Student-$t$ distribution with $\nu$ degrees of freedom. 

\begin{figure}[htbp]\centering
\includegraphics[width=\textwidth]{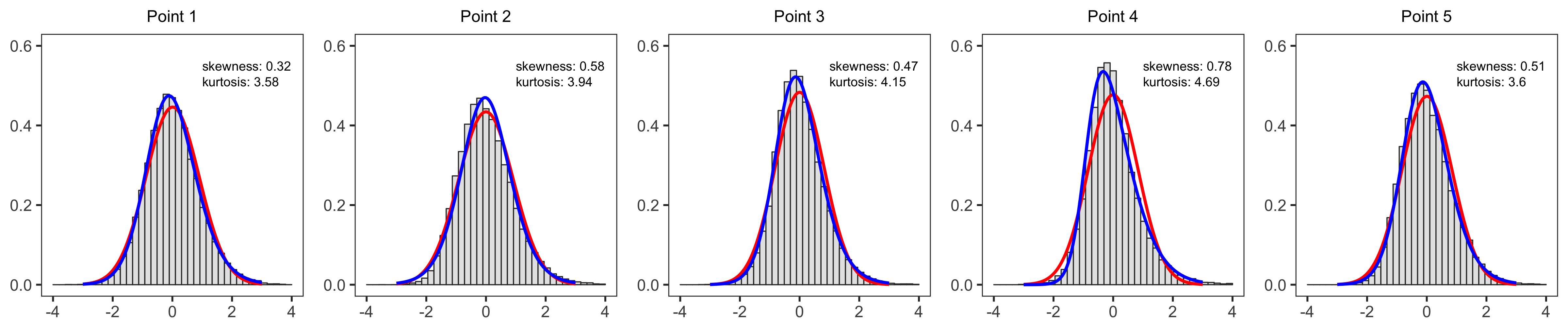}
\caption{Histograms of VAR(2) time series of the five points denoted in Fig.~\ref{fig:AnnCycle}. The marginal skew-$t$ distribution, obtained by marginalizing its multivariate parent, is denoted in blue. A reference Gaussian density has been added and denoted in red. The sample skewness and kurtosis are noted in each panel.}
\label{fig:PDF_comparison}
\end{figure}

Empirical justification for the model in \eqref{eq:Model} with non-Gaussian innovations is displayed in Fig.~3(b-c), where the pointwise estimates of skewness and kurtosis for the VAR(2) residuals are presented. Clear departures from zero and three, arise, respectively, in the southwest part of the country. Furthermore, in Fig.~\ref{fig:PDF_comparison} the marginal probability densities of the fitted multivariate skew-$t$ distributions for the five points denoted in Fig.~\ref{fig:AnnCycle} are contrasted with their respective histograms and a reference Gaussian distribution. The sample skewness is consistently positive across the various points, as is the excess kurtosis, in agreement with the values shown in Fig.~3(b-c).

\subsection{Inference}
One of the drawbacks of the parameterization used in \eqref{eq:densT}, called the \emph{direct parameterization} (DP), is that its parameters are not easy to interpret, as there is no simple relationship between them and the moments or cumulants of the distribution \citep{arellano2013centred}. This is most readily apparent in the location parameter $\bxi$, which bears no resemblance to typical centrality measures such as the mean or median. This is problematic also for inference, as the reductions in the parameter space that usually accompany assumptions such as a mean zero and a variance one do not materialize under this parameterization, as both the mean and the variance are functions of all of the above parameters (see Appendix). A centered parameterization (CP) based on the moments of the distribution was developed by \citet{azzalini1985class} for the DP of the skew-normal distribution; partly to offer a more intuitive parameterization, but more importantly, to respond to the singularity of its associated information matrix at $\balpha=\0$, which does not apply to the skew-$t$ distribution. Recently, \citet{arellano2013centred} extended this CP to the skew-$t$ distribution. In the univariate setting, with the skew-$t$ DP given by $(\xi,\omega^2,\alpha,\nu)$, the CP corresponds to the four centered moments $(\mu,\sigma^2, \gamma_1,\gamma_2)$, where $\gamma_1$ refers to the skewness coefficient and $\gamma_2$ is the coefficient of excess kurtosis. An implicit assumption in the CP is that these moments exist, requiring that $\nu > 4$, which may not hold in practice; and the exact value of $\nu$ imposes constraints on the feasible space of $\gamma_1$ and $\gamma_2$ (see Fig.~1 in \citet{arellano2013centred}). An examination of the marginals revealed that the values of the tuples $(\gamma_1,\gamma_2)$ encountered here were well within this feasible set. In the multivariate case, the first two components of the CP are the familiar mean vector and the covariance matrix, $\bmu$ and $\bSigma$, while the skewness term, $\bgamma_1$, consists of the vector of marginal skewness coefficients. The kurtosis term corresponds to the Mardia index of multivariate kurtosis (see Appendix), denoted by $\gamma_2^M$.

Because the singularity at $\balpha=\0$, which characterizes the skew-normal likelihood, is lacking here, maximization of the skew-$t$ likelihood is often the method of choice to conduct parameter inference. However, we find that when maximizing the log-likelihood to each set of residuals, the maximum likelihood estimator (MLE) tends to favor a more precise representation of the correlation structure at the expense of the marginal structure. This is not acceptable for our current application, which places a strong emphasis on the accurate representation of the marginal structure given the sensitivity of wind power density to wind speed. We considered adding a penalization term to the likelihood to constrain the optimization towards an improved marginal fit, but we obtained better results with a method of moments approach based on the above CP, as it afforded us greater control in isolating the fitting of the marginal structure from that of the spatial dependence. More specifically, using the CP gave us the possibility to directly impose the condition that $\bmu = \0$ and adopt a parametrized form for the correlation matrix $\hat{\bSigma}$ for each region, thereby considerably reducing the CP parameter space. $\hat{\bSigma}$ is parameterized by  $\mathcal{M}(\phi,\kappa)$, a Mat{\'e}rn correlation function with range parameter $\phi$ and smoothness $\kappa$. The latter was fixed at 1.5, and $\phi$ was fit by OLS. Good agreement was found  between the sample estimates of correlation of the VAR(2) residuals and the corresponding value from the OLS-estimated Mat{\'e}rn correlation function. Furthermore, errors from the OLS fit did not meaningfully contaminate the model-simulated seasonal variability in Section \ref{sec:saudi} (Fig.~S.10). Although the mapping from DP to CP is available in closed-form, the inverse map is not, and so it must be computed numerically. In fact, only the mapping $(\bgamma_1,\gamma_2^M) \mapsto (\bdelta,\nu)$ needs to be specified, since the rest of the DP parameterization follows directly from these parameters and those of the CP parameters (see sections 6.2.2 and 6.2.3 in \citet{azzalini2014skew} for details).  The \verb|sn| package \citep{azzalini2016r} implements an algorithm which seeks the $(d+1)$-vector $(\bdelta^\top,\nu)$ that minimizes the $\ell^2$-distance between $(\bgamma_1(\bdelta,\nu)^\top, \gamma_2^M(\bdelta,\nu))$ and its CP counterparts. As is common with many numerical optimization problems, finding a global maximum, or minimum in this case, is not guaranteed. In the few regions where such a situation arose, i.e., the $\ell^2$-distance does not vanish, the difference between the CP implied by the DP and the original CP was sufficiently small to be of no concern. 

The definition of the $\bgamma_1$ vector naturally suggests the use of sample estimates of the skewness coefficients at each gridpoint. For the multivariate kurtosis coefficient, $\gamma_2^M$, we considered its sample estimate, however, in some cases it yielded unsatisfactory results in terms of the implied skewness of the marginals once mapped to $\nu$.  In such cases, we replaced the sample estimate with the value which, together with $\hat{\bgamma_1}$, minimizes the $\ell^2$-distance between $\bgamma_1(\bdelta,\nu)$ and $\bgamma_2(\bdelta,\nu)$ and its sample counterparts, where here the dependence on $(\bdelta,\nu)$ indicates that these are the respective moment vectors after performing the mapping using the \verb|sn| package. 

As an indication of the sampling variability, we considered bootstrap standard errors based on 100 bootstrap samples for an arbitrarily chosen ensemble member and found that estimates of these for $\gamma_1$ across the domain were contained in $(0.014,0.035)$ and $\gamma_2$ in $(0.035, 0.179)$. Regarding the mapping from CP to DP, we recorded for each cluster the mean-squared-error (MSE) between the sample $\gamma_1$ and $\gamma_2$ of their constituent gridpoints and the respective values implied by the DP. Note that the errors corresponding to $\gamma_1$ serve as a reliable indicator of the accuracy of the mapping between CP and DP, whereas those for $\gamma_2$ are contaminated by an inherent characteristic of the DP paramerization, whose value of $\nu$ is inherited by all of the marginal skew-$t$ distributions. Therefore the latter errors also reflect the discrepancy between the actual marginal $\nu$ and this inherited value. The bootstrap estimates of the means and standard deviations for the MSE of $\gamma_1$, across the domain, were contained in $(0.002, 0.071)$ and $(0.01,0.095)$, respectively. The analogous values for the MSE of $\gamma_2$ are $(0.032,0.49)$ and $(0.02, 0.89)$.

\subsection{Model validation}\label{subsec:validation}

In order to assess the performance of the model in simulating realistic realizations of the daily surface wind speed over Saudi Arabia, the model was fit to three ensemble members, and 30 realizations were generated.  Several aspects of the simulated time series can be investigated as a basis for validation, such as the distribution of excursion durations above or below some fixed threshold as in \citet{ailliot2012markov}. Had wind direction been part of the analysis, for instance, the joint and marginal distribution of wind speed and direction, as well as their respective autocorrelation functions may have been investigated \citep[e.g.,][]{hering2015markov}. Given our interest in capturing low frequency aspects of wind speed, we focus here on the marginal distributions of daily wind speed and its serial dependence. However, for illustrative purposes, we have included an examination of the distribution of excursion durations in the supplementary material (Fig.~S.8 and associated section).

\begin{figure}[t]\centering
\includegraphics[width=\textwidth]{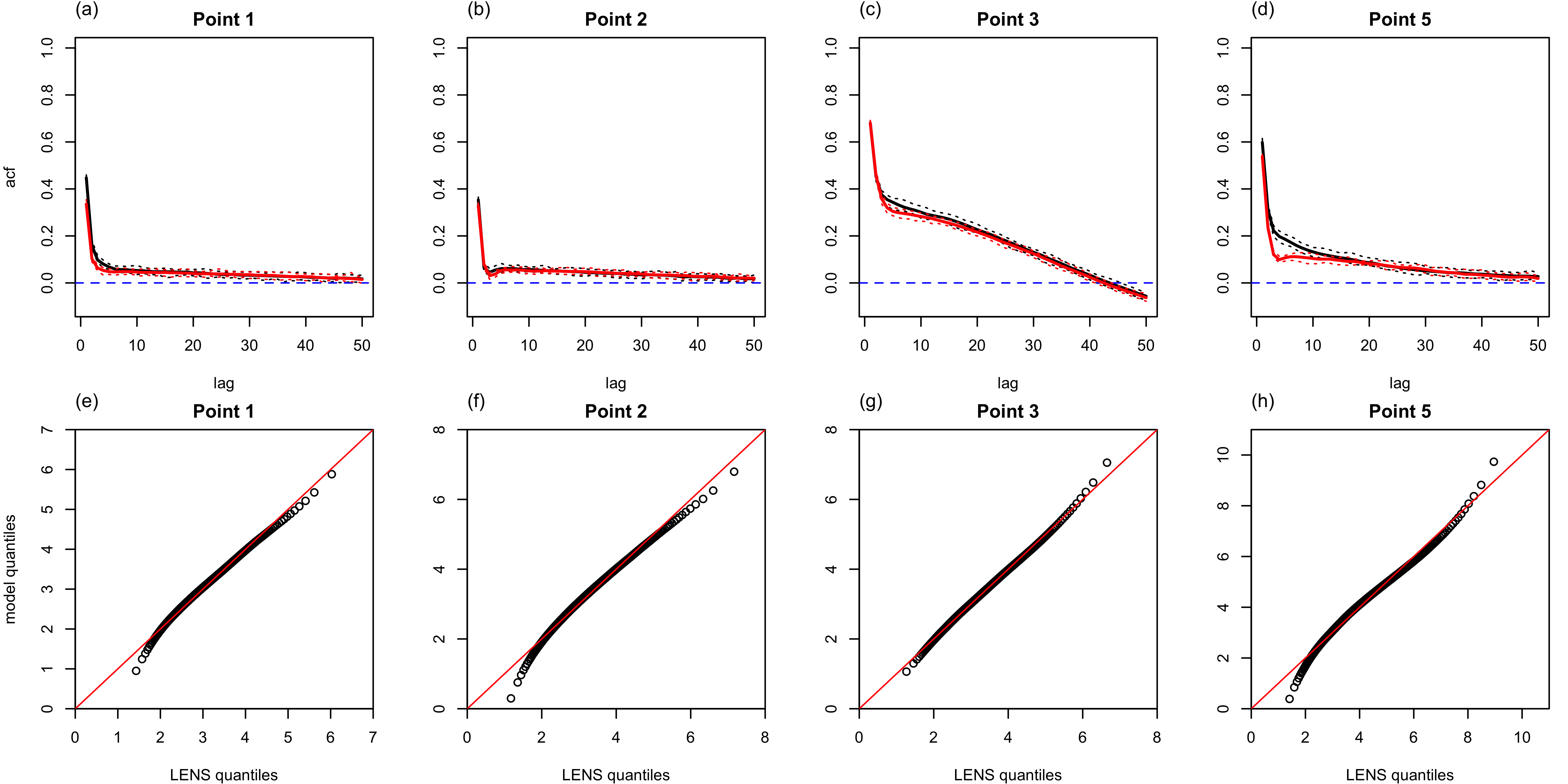}
\caption{(a-d) Average autocorrelation functions of 30 realizations of daily wind speed of (red) the proposed model and (black) the LENS, for Points 1, 2, 3 and 5 indicated in Fig.~\ref{fig:AnnCycle}. Dotted red and black lines represent the maximums and minimums of the simulated and LENS values, respectively, at each lag. A horizontal dashed blue line has been placed at zero. (e-h) Mean quantile-quantile plots of the daily surface wind speed for the LENS and the 30 model realizations, for the same points.}
\label{fig:SimACF}
\end{figure}

Figure \ref{fig:SimACF}(a-d) shows the mean autocorrelation function (ACF) for the 30 realizations and 30 LENS members, for four of the five representative points displayed in Fig.~\ref{fig:AnnCycle}, as the ACF for Point 4 is quite similar to that of Point 2, and so is excluded. The realizations adequately reproduce this feature of the LENS data, despite the marked differences among the autocorrelation structures across the various points. In particular, the ACFs in Points 1 and 2 drop sharply beyond the first few lags, whereas Point 3, which is in a region where both the northerly Shamal trade winds and southerly summer monsoon winds interact, shows a temporal dependence that remains relevant beyond the 50-day lag. 

A comparison of the agreement between wind speed distributions is displayed across Fig.~\ref{fig:SimACF}(e-h). The mean quantile-quantile (QQ) plots of the first three points show very good agreement between the simulated distributions and the LENS. Slight discrepancies emerge in the lower quantiles, most notably in Point 5 where even negative wind speeds are simulated, but this is a common drawback of models that do not have mechanisms in place to ensure non-negative values \citep{ailliot2012markov}. The discrepancies occur below the 2 m/s level, speeds at which common wind turbines are idle, therefore, these can be disregarded for the purposes of our application. Of greater importance is the upper tail of the QQ plots and the fit in all four cases is acceptable. 

\section{Wind Energy in Saudi Arabia}\label{sec:saudi}

The average wind speed or wind power density (WPD) at measurement height or adjusted-to-hub height is frequently used in assessments of wind energy potential. WPD refers to the wind power that is available per unit area, and is given by $\frac{1}{2} \rho w^3$, where $\rho$ represents the air density and $w$ is the wind speed. Several methods exist for extrapolating near surface wind measurements to heights at which wind turbines operate \citep{emeis2012wind}. Here we use the power law method, which assumes that the vertical wind profile at height $z$ is given by $w(z) = w(z_r) (z/z_r)^a$, where $z_r$ is a reference height and $a$ is the power law exponent, which depends on the surface roughness and the thermal stability of the boundary layer. Typically, $z_r$ represents the height at which observations are available. While the recommended value of $a$ varies according to the surface type, it is frequently fixed at $1/7$; this value has been found to be appropriate over open land surfaces, under neutral atmospheric stability. Since this value has been used in other wind studies in this region \citep[e.g.,][]{rehman2007wind} and the coarse resolution of the LENS dataset already limits the representativeness of our estimates, we use $a=1/7$ henceforth. 

Our computation of WPD starts with surface wind speed, defined in the LENS dataset as the wind speed at the lowest model layer (variable ZBOT in the database), which is then extrapolated to an 80-meter height using the above power law formula. The value of air density is assumed to be constant across space and time, and equal to 1.225 $kg\,m^{-3}$. Given the general increase in wind speeds during spring and summer, we focus on the March-April-May (MAM) and June-July-August (JJA) WPD seasonal averages, which are derived from daily WPD estimates; and we consider the last 30 years of the LENS dataset for comparability with recent studies. The mean MAM WPD averages approximately 36 $Wm^{-2}$ across Saudi Arabia, and it increases to 39 $Wm^{-2}$ during JJA, particularly along the northern coast of the Persian Gulf and west of the Asir Mountains, reflecting the strengthening of the northerly Shamal trade winds and the southerly monsoon winds, respectively (Fig.~S.9). 

\begin{figure}[!bh]\centering
\includegraphics[width=.7\textwidth]{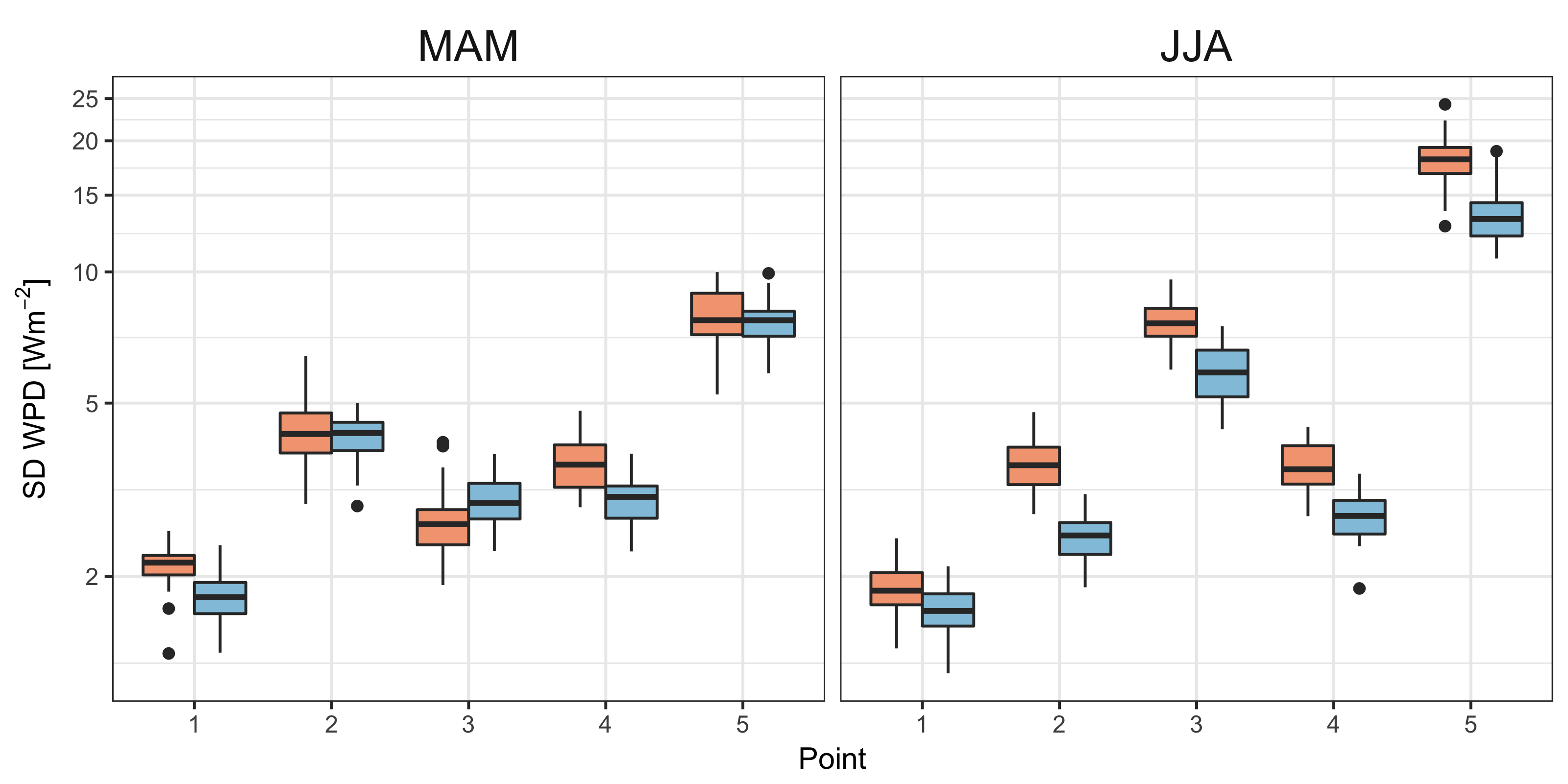}
\caption{Standard deviation of wind power density (WPD) at 80 meters for 1976-2005, in $Wm^{-2}$ for the five points denoted in Fig.~\ref{fig:AnnCycle}, during the (a) May-April-May (MAM) season and the (b) June-July-August (JJA) season. At each point, the orange boxplot corresponds to the WPD estimates derived from the LENS; the blue boxplots correspond to 30 generated realizations of the proposed model. Log10-scale is used on the y axis.}
\label{fig:WPDPoints}
\end{figure}

\begin{figure}[!hb]\centering
\includegraphics[width=\textwidth]{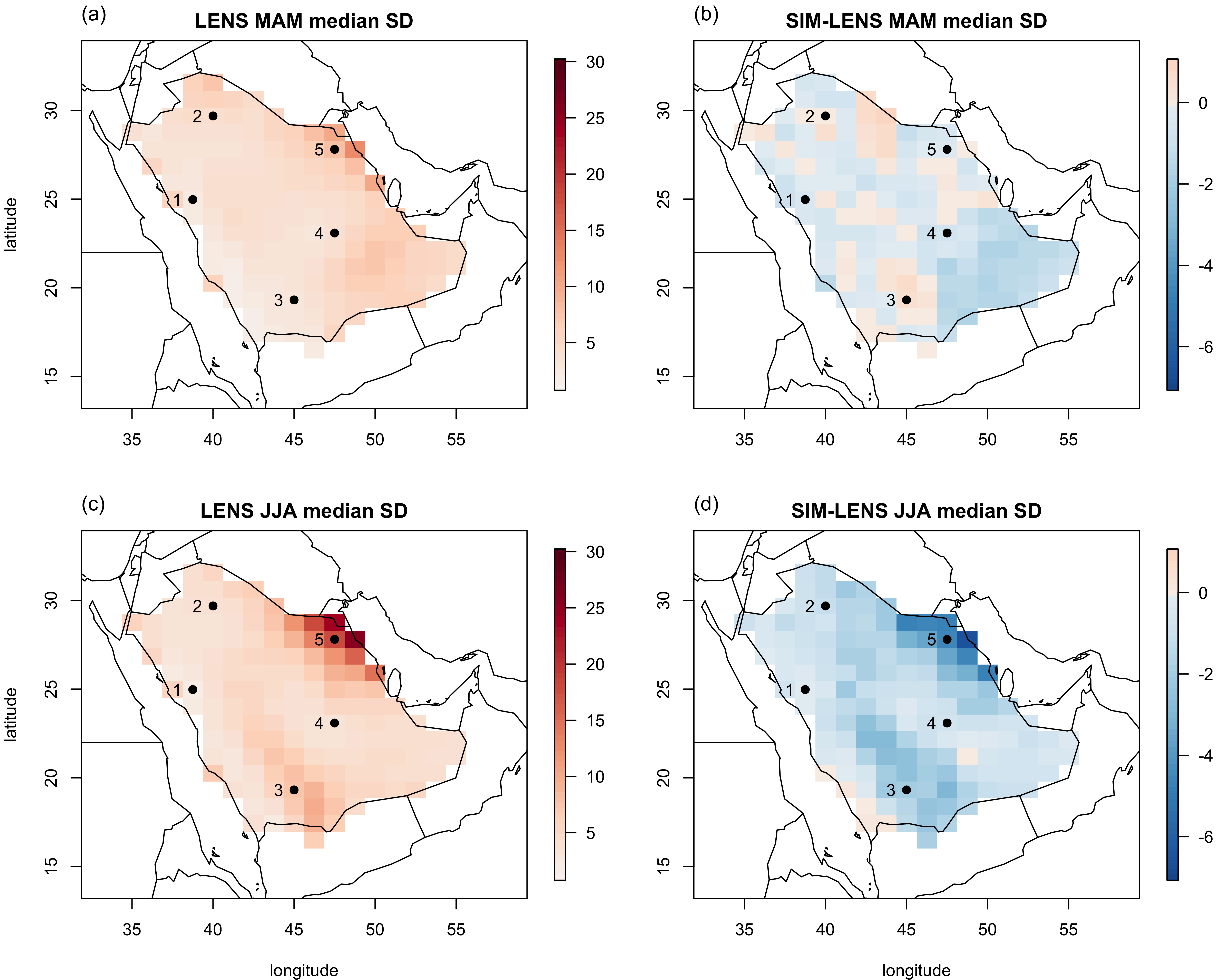}
\caption{Median standard deviation (SD) of 80 meter WPD, over 1976-2005, in $Wm^{-2}$ for the (a,c) LENS dataset over May-April-May (MAM) and June-July-August (JJA) and the (b,d) differences with respect to the estimates derived from the 30 generated realizations. Selected points denoted in Fig.~\ref{fig:AnnCycle} are also depicted.}
\label{fig:WPDMap}
\end{figure}

Figure \ref{fig:WPDPoints} displays the standard deviation (SD) of the LENS and simulated seasonal WPD estimates for both seasons. Median differences across seasons and points tend to follow the fluctuations of the annual cycles denoted in Section \ref{sec:data}, as higher wind speeds tend to be accompanied with greater variability. For instance, the median is nearly constant at Point 1, where the annual cycle experiences muted seasonal variation, but it more than doubles near the coastal areas of the Persian Gulf at Point 5, increasing from over 6 $Wm^{-2}$ in MAM to nearly 15 $Wm^{-2}$ in JJA. The fluctuations of the seasonal SD due to internal climate variability is shown to be considerable. At Point 5, the SD of JJA WPD varies between less than 15 and nearly 25 $Wm^{-2}$ across the 30 LENS realizations. At other points, the variation is smaller yet still relevant, as in Point 2 or 3 for MAM, where the highest value is nearly double that of the lowest. Comparing the values from the LENS with those implied by the model realizations suggests that the model reproduces the median and variability in seasonal WPD reasonably well during MAM, and this applies to the rest of the domain (Fig.~\ref{fig:WPDMap}(b)). However, negative biases arise in the median during JJA, especially at Point 5, where the bias reaches close to 4 $Wm^{-2}$. Figure \ref{fig:WPDMap}(d) shows how this bias also extends to the neighboring points along the Persian Gulf coastline. 

In JJA, the median SD (Fig.~\ref{fig:WPDMap}(c)), as represented in the LENS, displays a spatial pattern closely resembling that of the mean WPD shown in Fig.~S.9. However, the generated realizations have difficulty in capturing this rise in variability, as negative biases in excess of 4 and 2 $Wm^{-2}$ can be seen over the areas surrounding points 5 and 3, respectively. Analogous results apply to other quantiles, such as the 5\% and 95\% quantiles. To assess the impact of the errors of the OLS estimation on the WPD analysis, we fit the multivariate skew-$t$ distributions using the sample correlation matrices in place of the parameterized Mat\'ern correlation matrices, and we reported the results in Fig.~S.10.

\section{Discussion and Conclusions}\label{sec:disc&conc}

This study proposed a model for daily wind speeds at multiple sites based on a multivariate skew-$t$ distribution. Most recent models developed for wind speed and the 2-dimensional wind field combine a Hidden Markov Model (HMM), to account for the influence of distinct weather regimes, with an autoregressive structure to capture the high degree of autocorrelation that is common in atmospheric flow \citep{ailliot2006autoregressive,pinson2012adaptive,hering2015markov,bessac2016comparison}.  These Markov-Switching AutoRegressive (MS-AR) models have been shown to adequately represent the marginal distribution of daily wind speed, as the Markovian structure provides the flexibility necessary to capture its higher order moments. Recent efforts have attempted to extend this framework to the multisite setting; however, the number of sites is usually limited as inference quickly becomes problematic, in part due to the computationally demanding Expectation-Maximization (EM) algorithm that is used to deal with the latent Markov chain. Here we proposed a model that retains the vector-autoregressive structure, but that uses a neighbor-based scheme to reduce its dimensionality, and that replaces the latent weather regime process with a multivariate skew-$t$ distribution in the innovations to reproduce the observed skewness and excess kurtosis. The model was applied to a spatial domain of 149 points, which we partitioned into smaller regions of sizes ranging between 3 and 31 points where the assumption of stationarity is plausible, and we then performed the fit independently at each region. Our findings show that the model adequately matches the marginal distributions of daily wind speed, as well as the autocorrelation. However, the model shows some difficulty in matching the distribution of excursion durations above or below a high and low threshold, respectively, especially when the degree of persistence between the two excursion types is markedly different. We conducted an analogous assessment replacing the innovations in eq.~\eqref{eq:Model} with multivariate Gaussian distributions. As expected, the fit of marginal distributions showed a clear degradation, however, both the autocorrelation functions as well as the excursion distribution were nearly identical to those presented in Fig.~\ref{fig:SimACF}, see Figs S.11 and S.12. This highlights the relevance of the proposed mean structure to the performance in these two aspects, and suggests that adding seasonally varying vector-autoregressive coefficients may alleviate the shortcoming in the latter. Further work will involve incorporating a latent process to account for the large-scale variation of the wind field. Preliminary work reveals that such variation can be incorporated by exploiting the fact that the multivariate skew-$t$ distribution is constructed from a multivariate skew-normal distribution, which has the convenient property that sums of it with a normal distribution remains within the skew-normal family. 


The model was used to assess the sensitivity of wind resource potential in Saudi Arabia to the natural variability of the climate system. Studies devoted to the quantification of this resource have usually focused on the annual or seasonal mean wind speed,  its temporal variability or the energy it can potentially provide as measured by WPD. A recent work by \citet{yip2016wind} investigating the wind power potential of the Arabian Peninsula exceptionally considered other aspects, beyond its abundance, that are of particular relevance to the integration of wind energy into existing grid systems, such as the intermittency and persistence of the wind resource. To date, however, no study has addressed the impact of internal climate variability on wind energy related statistics, as its impact has only recently begun to be understood, thanks in part to projects such as the LENS that are designed to isolate its effects on the various earth system components. Our results, limited here to the variability of seasonal mean WPD based on 30 ensemble members from the LENS project, highlight the considerable effect of this source of variability on seasonal WPD. For instance, near the coastal areas of the Persian Gulf, the standard deviation of JJA WPD varies between 14 and 22 $Wm^{-2}$. At other points further inland the variation is not as severe, but is still relevant. Further work will expand on this analysis to include other aspects related to WPD, along the lines of \citet{yip2016wind}. The accuracy of these WPD estimates are limited by the coarse horizontal resolution of the LENS dataset, of approximately 1$^\circ$. In fact, we should expect to see departures from these as the resolution is refined, if only because of the better representation of the orography and the fine-scale processes that affect the boundary layer dynamics. A comparison of mean WPD at 80 meters (Fig.~S.9(c)) with the findings of \citet{yip2016wind}, based on a comparatively finer horizontal resolution of $0.5^\circ$ (latitude) and $0.67^\circ$ (longitude), hints at these potential discrepancies. Their work identifies the western mountains as offering more abundant wind resource than the Persian Gulf coastal areas, whereas here it is not the case. Ongoing work by the present authors with a high-resolution dataset are consistent with the findings of \citet{yip2016wind}; thus the estimates presented in this paper should be perceived as references from which to guide further inquiries performed at finer resolution.

The lack of reliable observational data over Saudi Arabia makes validation of the proposed model against observations challenging. A more structural hurdle to such a comparison is the difference in support between the LENS dataset, whose wind measurements at each grid cell correspond to area averages, and wind speed observations, obtained from meteorological masts or remote sensing systems such as LiDARs,  which may be regarded as point measurements. Overcoming this difficulty would involve the application of some form of downscaling technique which exceeds the scope of this work.

Ideally, climate centers should make projects such as the LENS regularly available to the research community at increasingly higher resolution to promote further analyses. However, projects of this magnitude, requiring scores of ensemble members, are costly from both a computational and a storage perspective. A cheaper alternative to running these numerical models is provided by stochastic weather generators, which attempt to reproduce the statistics of specific climate processes. The proposed model was fit to only three ensemble members, in order to assess its ability to reproduce the variability of the entire ensemble. Although some degree of underestimation was revealed over areas with abundant wind resource that warrants further study, it reasonably reproduced the variability of seasonal WPD, thus justifying its use as a potential stochastic generator of wind speed time series for other datasets.

\section*{Appendix: The Skew-$t$ Distribution}
In this appendix we collect a series of relevant properties and results related to the skew-$t$ distribution. 

If $\bY = \bxi + \bomega \bZ$, with $\bZ$ being the standard $ST_d$ variate, i.e., $\bZ \sim ST_d(\0,\bar{\bOmega},\balpha,\nu)$, then 
\begin{align*}
\bmu  & = \text{E}(\bY) = \bxi + \bomega \bmu_z, \quad  \nu > 1, \\
\bSigma  & = \text{var}(\bY) = \frac{\nu}{\nu -2} \bOmega - \bomega \bmu_z \bmu_z^\top \bomega, \quad \nu > 2,
\end{align*}
with $\bmu_z = b_\nu \bdelta$, where
\begin{equation*}
b_\nu = \frac{\sqrt{\nu} \, \Gamma \left\{\frac{1}{2} (\nu-1) \right\} }{\sqrt{\pi} \, \Gamma \left(\frac{1}{2}\nu \right)}, \; \nu > 1, \quad \text{and} \quad \bdelta = \left( 1 +\balpha^\top \bar{\bOmega} \balpha \right)^{-1/2} \bar{\bOmega} \balpha.
\end{equation*}

Mardia's measure of multivariate kurtosis \citep{mardia1970measures}, for a $d$-dimensional random variable $\bX$, is given by
\[  \gamma_2^M  = \text{E} \left[ \{(\bX-\bmu)^\top \bSigma^{-1} (\bX - \bmu)\}^2 \right] - d(d+2) \]
and for a $ST_d$ variate, assuming that $\nu > 4$,
\[ \gamma_2^M  = \frac{2d (d+2)}{\nu - 4} + \frac{4(d+2)}{(\nu-3)(\nu - 4)} \beta_0^2 + 
                          2 \left\{  \frac{2 \nu}{(\nu - 3)b_\nu^2}  - \frac{3 (\nu-3)^2-6}{(\nu-3)(\nu - 4)} \right\}\beta_0^4 \]
where $\beta_0^2 = \bmu_0^\top \bSigma^{-1} \bmu_0$, $\bmu_0 = \bomega \bmu_z$.

\bibliographystyle{asa}
\bibliography{JTSA}

\end{document}